\documentclass[runningheads]{llncs}
\usepackage[T1]{fontenc}
\usepackage{cite}
\usepackage{tikz}
\usetikzlibrary{positioning}
\usetikzlibrary {arrows.meta}
\usetikzlibrary{backgrounds}
\usepackage{amsmath,amssymb,amsfonts}
\usepackage{graphicx}
\usepackage{textcomp}
\usepackage{lmodern}
\usepackage{verbatim}
\usepackage[hidelinks]{hyperref}
\usepackage{orcidlink}
\usepackage{url}
\usepackage{xspace}
\usepackage{float}
\usepackage{wrapfig}
\usepackage[frozencache,cachedir=minted-cache]{minted}
\usepackage{varwidth}
\usepackage{subcaption}
\usepackage{hwemoji}

\usepackage{ifthen}

\newboolean{showcomments}
\setboolean{showcomments}{false}

\makeatletter
\newcommand{\mynote}[3]{%
  \ifthenelse{\boolean{showcomments}}{%
   \fbox{\bfseries\sffamily\scriptsize#1}%
   {\small\textsf{\emph{\color{#3}{#2}}}}}%
  {%
   \@bsphack
   \@esphack
  }%
}
\makeatother

\definecolor{asparagus}{rgb}{0.53, 0.66, 0.42}

\newcommand{\out}[1]{}

\definecolor{Cerulean}{rgb}{0.20,0.31,0.0.90}
\definecolor{RedOrange}{rgb}{0.70,0.59,0.0.20}
\definecolor{SomeGreen}{rgb}{0.1,0.60,0.1}
\definecolor{SomeRed}{rgb}{0.80,0.3,0.0.20}
\definecolor{lightGray}{rgb}{0.85,0.85,0.86} 
\definecolor{lGray}{rgb}{0.31,0.31,0.33} 
\definecolor{lYellow}{rgb}{0.99,0.78,0.07} 

\definecolor{SomeGreen1}{rgb}{0.3,0.80,0.3}
\definecolor{lBlue}{rgb}{0.30, 0.45, 0.66}
\newcommand{\indist}{\ensuremath{\approx}\xspace}
\newcommand{\pindist}{\ensuremath{\approx_0}\xspace}
\newcommand{\sigprot}{\texttt{SigProt}\xspace}
\newcommand{\attprot}{\texttt{AttProt}\xspace}

\newcommand{\ra}{\text{RA}\xspace}

\newcommand{\ssp}{\text{SSP}\xspace}
\newcommand{\ssprove}{\text{SSProve}\xspace}
\newcommand{\coq}{\text{Rocq Prover}\xspace}

\newcommand{\msg}{\ensuremath{m\xspace}}

\newcommand{\sig}{\ensuremath{\sigma\xspace}}
\newcommand{\A}{\ensuremath{\mathcal{A}\xspace}}

\newcommand{\cget}{\mintinline{Coq}{get}\xspace}
\newcommand{\cput}{\mintinline{Coq}{put}\xspace}
\newcommand{\cassert}{\mintinline{Coq}{assert}\xspace}

\newcommand{\csample}{\mintinline{Coq}{<$}\xspace}

\newcommand{\pkeygen}{\mintinline{Coq}{KeyGen}\xspace}
\newcommand{\psigprim}{\mintinline{Coq}{SigPrim}\xspace}
\newcommand{\psigprot}{\mintinline{Coq}{SigProt}\xspace}

\newcommand{\pprot}{\mintinline{Coq}{Prot}\xspace}
\newcommand{\pattprot}{\mintinline{Coq}{AttProt}\xspace}
\newcommand{\pattprim}{\mintinline{Coq}{AttPrim}\xspace}

\newcommand{\psigprimattR}{\mintinline{Coq}{SigPrimAtt}$_{\real}$\xspace}
\newcommand{\psigprimattId}{\mintinline{Coq}{SigPrimAtt}$_{\ideal}$\xspace}
\newcommand{\psigprimId}{\mintinline{Coq}{SigPrim}$_{\ideal}$\xspace}
\newcommand{\pattprimsig}{\mintinline{Coq}{AttPrimSig}$_{\text{\real}}$\xspace}
\newcommand{\pattprimsigId}{\mintinline{Coq}{AttPrimSig}$_{\text{\ideal}}$\xspace}
\newcommand{\psigprimR}{\mintinline{Coq}{SigPrim}$_{\text{\real}}$\xspace}
\newcommand{\pattprimR}{\mintinline{Coq}{AttPrim}$_{\text{\real}}$\xspace}
\newcommand{\pattprimId}{\mintinline{Coq}{AttPrim}$_{\text{\ideal}}$\xspace}

\newcommand{\egetpk}{\mintinline{Coq}{get_pk}\xspace}
\newcommand{\esign}{\mintinline{Coq}{sign}\xspace}
\newcommand{\eversig}{\mintinline{Coq}{ver_sig}\xspace}
\newcommand{\egetpka}{\mintinline{Coq}{get_pka}$^{\text{a}}$\xspace}
\newcommand{\eattest}{\mintinline{Coq}{attest}\xspace}

\newcommand{\everifya}{\mintinline{Coq}{verify}$^{\text{a}}$\xspace}
\newcommand{\ekgen}{\mintinline{Coq}{key_gen}\xspace}

\newcommand{\lseckey}{\mintinline{Coq}{SecKey}\xspace}
\newcommand{\lpubkey}{\mintinline{Coq}{PubKey}\xspace}

\newcommand{\ststate}{\mintinline{coq}{State}\xspace}
\newcommand{\hignore}{\mintinline{Coq}{heap_ignore}\xspace}
\newcommand{\ztosig}{\mintinline{Coq}{Z_to_SIG}\xspace}

\newcommand{\real}{\textcolor{lGray}{real}\xspace}
\newcommand{\ideal}{\textcolor{lYellow}{ideal}\xspace}

\newcommand{\kimport}{\textbf{\texttt{\#import}}}
\newcommand{\kput}{\textbf{\texttt{\#put}}}
\newcommand{\kget}{\textbf{\texttt{\#get}}}
\newcommand{\kret}{\textbf{\texttt{\#ret}}}

\def\checkmark{\tikz\fill[scale=0.4](0,.35) -- (.25,0) -- (1,.7) -- (.25,.15) -- cycle;}

\newcommand{\icoq}[1]{\mintinline{Coq}{#1}}

\newtheorem{game}{Game}
\newtheorem{hypothesis}{Hypothesis}

\newcommand\spacer[1]{\rule[-#1]{0pt}{#1}}

\def\BibTeX{{\rm B\kern-.05em{\sc i\kern-.025em b}\kern-.08em
		T\kern-.1667em\lower.7ex\hbox{E}\kern-.125emX}}

\begin{document}

\title{Formally-verified Security against Forgery \\of Remote Attestation using \ssprove}
\titlerunning{Security of Remote Attestation using \ssprove}
\author{Sara Zain\inst{1}\orcidlink{https://orcid.org/0009-0007-1533-5670} \and
Jannik M\"ahn \inst{2} \and 
Stefan K\"opsell \inst{1}\orcidlink{0000-0002-0466-562X} \and
Sebastian Ertel \inst{1}\orcidlink{https://orcid.org/0009-0000-3953-9810} }
\authorrunning{S. Zain et al.}
%
\institute{Barkhausen Institut, Dresden, Germany \\
\email{\{sara.zain, stefan.koepsell, sebastian.ertel\}@barkhauseninstitut.org} \\
\and \email{jannik.maehn@gmail.com}}

	\maketitle

	\begin{abstract}
%

    %

%
%
%
%
%
%

  %
  %

%
Remote attestation (\ra) is the foundation for 
trusted execution environments in the cloud and 
trusted device driver onboarding in operating systems.
However, \ra misses a rigorous mechanized definition of its
security properties in one of the strongest models, i.e.,
the semantic model. 
%
  %
  %
%
Such a mechanization requires 
the concept of State-Separating Proofs (SSP).
However, SSP was only recently implemented as a foundational framework in the \coq. 
Based on this framework, this paper presents the first mechanized formalization of the fundamental security properties of \ra. 
Our \coq development first defines digital signatures and
formally verifies security against forgery in the strong
existential attack model.
Based on these results, we define \ra and reduce the security
of \ra to the security of digital signatures.

  %
  %

%
Our development provides evidence that the \ra protocol
is secure against forgery.
Additionally, we extend our reasoning to the primitives
of \ra and reduce their security to the security of the
primitives of the digital signatures.
%
%
%


\end{abstract}



\section{Introduction}\label{sec:intro}
Remote attestation (\ra) is a fundamental security protocol for 
establishing authenticity in many digital systems.
It is a core building block in systems ranging from embedded 
IoT devices~\cite{10.1145/2592798.2592824,10.1145/2897937.2898083,10.1145/3654661} 
to cloud-based trusted execution environments~\cite{10.1145/3319535.3354216,10.1145/3319535.3354220}. 
However, despite its central role in establishing trust, \ra protocols are often 
specified without formal cryptographic models, 
and their security properties—particularly 
semantic security—remain informally stated or entirely unproven.

The Trusted Platform Module (TPM) specification exemplifies this issue~\cite{tpm-spec}.
While TPM specification, for example, outlines key generation and digital signature primitives 
and defines an attestation flow, it does not formalize the cryptographic properties it relies on, 
such as semantic security or unforgeability under chosen-message attacks. 
It also lacks a formal model of the adversary or a clear representation of 
what it means for attestation to be indistinguishable from an ideal execution.
Existing formal approaches for \ra often focus on functional correctness or 
adopt a Dolev-Yao model to show \emph{symbolic} security~\cite{pornin2005digital} 
but cannot establish \emph{stronger} semantic security. 
These models can verify correctness but assume 
cryptographic primitives such as encryption and decryption are correct. 
VRASED instantiates a hybrid (HW/SW) \ra co-design for embedded devices 
but relies on functional correctness~\cite{vrased}.
To prove semantic security is particularly challenging, 
and appropriate reasoning frameworks were only established recently~\cite{9919671}.
A novel methodology called State-Separating Proofs (\ssp)%
\footnote{
    While we assume familiarity with foundational frameworks like SSProve and 
    State-Separating Proofs, we summarize the key modeling elements 
    (packages, procedures, state memory) in Section~\ref{sec:TheoryFound} and Figure~\ref{fig:package} 
    to support accessibility.
}.
allows reasoning in a modular way about semantic security~\cite{ssp}.
Frameworks for cryptographic reasoning like CertiCrypt~\cite{barthe2009formal}
and (its extension) EasyCrypt~\cite{barthe2012easycrypt} 
have existed for a decade but remained hard to apply. 

We provide the first formal proof of semantic security for \ra protocols using \ssprove.
\ssprove{~\cite{ssprove}} is a \coq\footnote{Previously named as Coq} based framework for game-based cryptographic proofs, 
and reasonings are done in probabilistic relational Hoare logic (pRHL)~\cite{barthe2009formal}. 
We favored \ssprove because it provides access to the rich mathematical components library~\cite{mahboubi2021mathematical}.
We present the generalization of \emph{strong unforgeability} to 
\emph{perfect indistinguishability} to show that it reduces the security of 
\ra to the security of secure signatures in Section~\ref{sec:TheoryFound}. 
In Section~\ref{sec:vra}, the \ra protocol specifies correctness and indistinguishability properties, 
and prove the indistinguishability of attestation responses 
under the chosen-message attack. 
Our formalization includes an instantiation with RSA signatures, 
for which we implement and verify key generation (Section~\ref{sec:rsa}), 
and we ensure that all cryptographic assumptions are captured and mechanized.

Our contribution is publicly availble\footnote{https://github.com/Barkhausen-Institut/vRATLS/tree/main/theories/examples}. 
%
%

	\section{From Unforgeability to Indistinguishability}\label{sec:TheoryFound}
%
%
%
\vspace{-0.5em}
This section presents the security guarantees of digital signatures, 
focusing on the transition from strong existential unforgeability (sEUF-CMA) 
to indistinguishability. 
We will show how these properties are crucial for \ra and 
are framed within the \ssprove methodology, paving the way for a better 
understanding of \ra security.
Then, we show how these properties are recast within the \ssprove methodology through semantic indistinguishability games. 

We begin by recalling the standard textbook definitions of a digital signature scheme and its correctness~\cite{joy,companion}, 
followed by their unforgeability guarantees. 
Any reference to ``textbook definitions" will refer to these references throughout the rest of this paper. 
\vspace{-0.5em}
\subsection{Digital Signature Schemes}
A digital signature scheme 
$\Sigma =(\text{KeyGen},\text{Sign},\text{VerSig})$ 
defined over a message space $\mathcal{M}$ consists
of three probabilistic polynomial-time algorithms:
\vspace{-0.5em}
\begin{itemize}
    \item \textbf{KeyGen} generates
    a pair of keys $(sk, pk)$, where $sk$ is the secret signing
    key and $pk$ is the public verification key. 
    \item\textbf{Sign} takes the secret signing key $sk$ and
    a message $m\in\mathcal{M}$ to generate a signature $\sigma$ on the message.
    \item\textbf{VerSig} uses the public key $pk$ to verify whether
    a signature $\sigma$ was generated from a message $m$.
\end{itemize}
\vspace{-0.5em}
Functional correctness ensures that the verification process
succeeds for legitimate signatures.
\vspace{-0.5em}
\begin{definition}[Functional Correctness]\label{def:sig:correct}
  A signature scheme $\Sigma$ is correct if, for all $ m\in\mathcal{M} $ 
  and all key pairs ($sk, pk$) generated by \textbf{KeyGen}, we have:
  \[ \textbf{VerSig}(pk, m, \textbf{Sign}(sk,m)) = true. \]
\end{definition}
\vspace{-0.5em}
\vspace{-0.4em}
\subsection{Strong Unforgeability}
Here, we recall the notion of unforgeability in a digital signature scheme. 
There are two key notions in digital signature. 
The first is existential unforgeability under chosen-message attacks (EUF-CMA), where 
an adversary has to forge a signature of any \emph{new} message, i.e., 
a message without a previously generated signature~\cite{goldwasser1988digital}.
The second is a stricter variant, i.e., strong existential unforgeability (sEUF-CMA) (our research focus), where
the adversary has to forge a \emph{new 
signature} of any message, i.e., including messages with a previously generated signature.
The formal definition of sEUF-CMA is as follows: 
\vspace{-0.5em}
\begin{definition}[sEUF-CMA]\label{def:seuf-cma}
  A digital signature scheme $\Sigma$ is strongly
  existentially unforgeable	under a chosen-message attack if, for any
  probabilistic polynomial-time adversary $\mathcal{A}$,
  the probability of forging a new, valid signature for any message—including previously 
  signed messages—is negligible. 
  The advantage of $\mathcal{A}$ in breaking the sEUF-CMA security is
  \[
  \text{Adv}_{\Sigma}^{\text{sEUF-CMA}} (\mathcal{A}) :=
  \text{Pr}\big[ \text{Exp}_{(\Sigma,\mathcal{A})}^{\text{sEUF-CMA}}()\big]< \text{negl}
  \]
  where the $\text{Exp}_{(\Sigma,\mathcal{A})}^{\text{sEUF-CMA}}$ is defined in Game~\ref{game:seuf-cma}. 
\end{definition}
\vspace{-0.5em}
In the sEUF-CMA experiment (Game~\ref{game:seuf-cma}), 
the challenger provides access to a signing oracle while 
preventing trivial wins. 
The strong unforgeability condition rules out signature 
re-randomization or malleability, which is especially critical in \ra
settings where replay or signature variance can lead to security breaches.
\vspace{-1.2em}
  \begin{game}\label{game:seuf-cma}
    \texttt{\underline{$\text{Exp}_{(\Sigma, \mathcal{A})}^{\text{sEUF-CMA}}( )$}}
    \begin{itemize}
      \item The \emph{challenger} generates a key pair $(sk,pk)$ using KeyGen$( )$ and 
      initializes a set $S$ to keep track of signed message-signature pairs.
      \item Adversary $\mathcal{A}$ is given access to a signing oracle and the public key $(pk)$. 
      It may query the oracle for arbitrary messages.
      \item For each query $m_i$, the \emph{challenger} computes $\sigma_i \leftarrow Sign(sk, m_i)$, 
      return $\sigma_i$ to $\mathcal{A}$, and stores $(m_i, \sigma_i)$ in the set $S$.
      \item Finally, A outputs a candidate forgery $(m^*, \sigma^*)$. It wins if:
      \begin{enumerate}
        \item $VerSig(pk, m^*, \sigma^*) = true $, and
        \item $(m^* , \sigma^*) \notin S $, i.e., this exact signature was not returned by the oracle.
      \end{enumerate} 
    \end{itemize}
  \end{game}
\vspace{-0.9em}
\subsection{Toward Indistinguishability}
\vspace{-0.5em}
%
In \ra, adversaries may interact repeatedly with the signing oracle. 
We know that sEUF-CMA guarantees that adversaries cannot forge fresh signatures or produce 
new valid signatures on known messages. 
Now, we need to ensure that all signing executions appear indistinguishable. 
That is, the internal behavior of the signing process—such as randomness usage, timing, 
or subtle variations, should not leak information.
Even if forgery remains impossible, the ability to distinguish 
between different signing behaviors should not threaten \ra security.

To address this, we adopt \emph{semantic indistinguishability}: 
an adversary must fail to forge and be unable to distinguish 
whether it interacts with a \emph{real} signer or an \emph{idealized} one that leaks no 
information beyond valid signatures. 
We formalize this as an indistinguishability game (Figure~\ref{fig:real-idealGame}): 
the adversary submits signing and verification queries and must guess 
whether it interacts with the \emph{real} implementation or a simulator enforcing \emph{ideal} behavior.
Winning this game must be computationally infeasible.
Our formalization is based on game-based proofs~\cite{shoup2004sequences}.
It is an approach to proving cryptographic properties, such as correctness, 
indistinguishability, and unforgeability, through a sequence of game transitions, 
with reasoning conducted in probabilistic relational Hoare logic (pRHL)~\cite{barthe2009formal}. 
Each transition ensures that no adversary can distinguish real from ideal executions.
We use \ssprove, a \coq-based framework for machine-checked cryptographic proofs, 
to express and verify the guarantees of indistinguishability. 
\ssprove builds on the State-Separating Proof (SSP) methodology, 
which structures cryptographic reasoning around packages of procedures
that operate over a shared state~\cite{ssp}.
\paragraph{Packages and Procedures.}
Readers familiar with SSP can safely skip this paragraph.
A package in \ssprove defines a set of procedures operating over a common state. 
For example, in Figure~\ref{fig:package}, package $\mathcal{E}$ 
may define procedure \icoq{z}, which depends on imported procedure types \icoq{x} and \icoq{y}. 
%
\begin{figure}
  \centering
  \begin{tikzpicture}[
      scale=0.75, transform shape
    ]
		\node[] (P) at (0,0)  {
      $\mathcal{E}$
    };
    \node[] (c) [below = 0.5cm of P] {
      \begin{minipage}{0.30\columnwidth}
      \begin{minted}[fontsize=\scriptsize,escapeinside=&&]{coq}
def Z:
  let z := X () + Y () in
  #put @&$\mathcal{S}$& z ;;
  z <- #get @&$\mathcal{S}$& ;;
  n <$ &$\mathbb{N}_6$& ;;
  #ret (z + n)
        \end{minted}
      \end{minipage}
    };
    \node[] (s) [above right = 0cm and 0cm of c.north west] {
      $\mathcal{S}$\mintinline{Coq}{: nat}
    };
   \draw[->] ([xshift=-2cm,yshift=-0.5cm]c.north west) --
   ([xshift=-0.2cm,yshift=-0.5cm]c.north west);
    \node[] (X) [above left = -0.5cm and 0.5cm of c.north west] {
      \begin{minipage}{0.2\columnwidth}
        \begin{minted}[fontsize=\scriptsize]{coq}
X: unit -> nat
        \end{minted}
      \end{minipage}
    };

    \draw[->] ([xshift=-2cm,yshift=0.5cm]c.south west) --
    ([xshift=-0.2cm,yshift=0.5cm]c.south west);
    \node[] (Y) [above left = 0.5cm and 0.5cm of c.south west] {
        \begin{minipage}{0.2\columnwidth}
          \begin{minted}[fontsize=\scriptsize]{coq}
Y: unit -> nat
          \end{minted}
        \end{minipage}
      };

    \draw[->] ([xshift=0.2cm]c.east) -- ([xshift=1.5cm]c.east);
    \node[] (Y) [above right = 0cm and 0.5cm of c.east] {
      \mintinline{Coq}{Z:nat}
    };

    \node[] () [left = 2.3cm of P] { Imports };
    \node[] () [right = 2.3cm of P] { Exports };

    \draw[] (c.north west) -- (c.north east);
    \draw[] (c.north west) |- (s.north) -| (c.north east);
    \draw[] (s.north west) |- (P.north) -| (c.north east);
    \draw[] (c.north east) |- (c.south) -| (c.north west);

	\end{tikzpicture}
	\caption{A package $\mathcal{E}$ in \ssprove.}
  \label{fig:package}
\end{figure}
%
\ssprove supports sequential composition ($\mathcal{E}_1 \circ \mathcal{E}_2$) by inlining procedures, 
provided the interface matches and states align.
Procedures are written using a state monad code to handle randomness and memory~\cite{10.1145/143165.143169}. 
Commands such as \cget, \cput, \cassert, and \csample 
(for sampling) allow precise control over state and randomness. 
All procedures are annotated using (\texttt{\#}), and memory locations (e.g., \icoq{@}$\mathcal{S}$) abstract away implementation details. 
%
A package can define an interface, 
such as signing and verification procedures, and a stateful implementation

In \ssprove, we model the signature process as two distinct packages: 
one that implements the \emph{real} functionality and the other that provides an \emph{idealized}, 
abstracted view of signatures (Figure~\ref{fig:real-idealGame}).
\vspace{-1.2em}
\begin{figure}
  \centering
  \begin{tikzpicture}[
      scale=0.75, transform shape
    ]
	\node[] (real) at (0,0)  {
		$\mathcal{G}_{\text{\real}}^{\Sigma}$
	};
	\node[] (code-r) [below = 0cm of real] {
		\begin{minipage}{0.25\columnwidth}
			\begin{minted}[fontsize=\scriptsize,escapeinside=@@,autogobble]{coq}
@\textcolor{white}{empty line}@
(@$sk$@,@$pk$@) @$\leftarrow$@ @$\Sigma.$@KeyGen()

get_pk()
  return @$pk$@

sign(@$m$@)
  @$\sig\;:=\;$@ @$\Sigma.$@Sign(@$sk,\msg$@)
  
  return @$\sig$@

ver_sig(@$\msg,\sig$@)
@$b:=$@ @$\Sigma.$@VerSig(@$pk,\msg,\sig$@)
  return @$b$@
			\end{minted}
		\end{minipage}
	};

	\draw[] (code-r.north west) |- (real.north)   -| (code-r.north east);
	\draw[] (code-r.north west) |- (real.south)   -| (code-r.north east);
	\draw[] (code-r.north east) |- (code-r.south) -| (code-r.north west);

	
	\node[] (ideal) [right = 2.5cm of real]  {
		$\mathcal{G}_{\text{\ideal}}^{\Sigma}$
	};
	\node[] (code-i) [below = 0cm of ideal] {
		\begin{minipage}{0.25\columnwidth}
			\begin{minted}[fontsize=\scriptsize,escapeinside=@@,autogobble,stripnl=false,highlightlines={1,9,13},highlightcolor=lYellow!20]{coq}
@$S\;=\emptyset$@
(@$sk$@,@$pk$@) @$\leftarrow$@ @$\Sigma.$@KeyGen()

get_pk()
  return @$pk$@

sign(@$m$@)
  @$\sig\;:=\;$@ @$\Sigma.$@Sign(@$sk,\msg$@)
  @$S\;:=\;S\;\cup\;(\msg,\sig)$@
  return @$\sig$@

ver_sig(@$\msg,\sig$@)
  @$b:=$@ (@$\msg,\sig$@) @$\in$@ @$S$@
  return @$b$@
			\end{minted}
		\end{minipage}
	};

	\draw[] (code-i.north west) |- (ideal.north)  -| (code-i.north east);
	\draw[] (code-i.north west) |- (ideal.south)  -| (code-i.north east);
	\draw[] (code-i.north east) |- (code-i.south) -| (code-i.north west);

	\node[] (adv) [below right = 5.2cm and 1cm of real]  {
		$\mathcal{A}$ 
	};

	\draw[->] (adv.north) -- ([yshift=0.2cm]adv.north) -| ([xshift=-1.4cm,yshift=0.2cm]adv.north) -| ([yshift=-0.1cm]code-r.south);
	\draw[->] (adv.north) -- ([yshift=0.2cm]adv.north) -| ([xshift=1.4cm,yshift=0.2cm]adv.north) -| ([yshift=-0.1cm]code-i.south);

	\node[] (win) [above = 0.4cm of adv]  {
		 real\hspace{0.7cm}🏆\hspace{0.6cm}ideal
	};

\end{tikzpicture}
  \caption{ Game-Pair for indistinguishability 
  ($ \mathcal{G}_{\text{ \textcolor{lGray}{R} \textcolor{lYellow}{I}} }^{\Sigma}$) 
  }
  \label{fig:real-idealGame}
\end{figure}
\vspace{-0.9em}
The adversary interacts with one of these packages—it can submit 
signing queries and attempting to verify signatures—and 
must distinguish whether 
it interacts with the real or the ideal implementation. 
%
%
Proving that these packages are indistinguishable ensures that signing interactions leak no information beyond what correctness requires.

The \emph{real game} models the real (honest) signature scheme: 
the adversary may query the signing oracle and later submit a message-signature pair for verification. 
If the signature is valid, the adversary wins. 
In the \emph{ideal game}, a simulator responds to signing queries 
and records issued signatures in a set $S$. 
When the adversary attempts to verify a signature, 
the simulator only accepts message-signature pairs that were previously 
output by the signing oracle. Any attempt to verify 
an unissued (potentially forged) signature automatically fails. 
This technique ensures that guessing a valid signature in 
the ideal game is infeasible—not because of randomness, 
but because the simulator enforces a strict policy: only known, 
previously issued signatures succeed.

The proof of indistinguishability ensures that such forgeries are equally 
infeasible in the real game—if the two games are indistinguishable, then no such exploit exists.
Formally, we express this guarantee using perfect indistinguishability:
\vspace{-0.5em}
\begin{definition}[Perfect Indistinguishability ($\approx_0$)~\cite{ssprove,joy,companion}]
  Let $\mathcal{G}_1$ and $\mathcal{G}_2$ be two games with the same interface. 
  They are perfectly indistinguishable, denoted $\mathcal{G}_1\approx_0\mathcal{G}_2$,
  if for all probabilistic polynomial-time (PPT) adversaries $\mathcal{A}$,
  $$\text{Adv}_{(\mathcal{G}_1, \mathcal{G}_2)} (\mathcal{A}) 
  = |\;\text{Pr}\big[\mathcal{A}\circ\mathcal{G}_{1}\big]-\text{Pr}\big[\mathcal{A}\circ\mathcal{G}_{2}\big]\;| = 0 $$
\end{definition}
\vspace{-0.5em}
The $\mathcal{A}\circ\mathcal{G}$ denotes the adversary interacting with game $\mathcal{G}$.
Perfect indistinguishability implies that no adversary—regardless of its 
strategy—can gain any advantage in distinguishing between the real and ideal systems.
Unlike indistinguishability, which only needs the advantage to be negligible, 
perfect indistinguishability ensures exact equality to zero. 

In our \ssprove development setting, the real and ideal games differ only in their treatment of signature verification. 
The ideal game maintains an explicit set $S$ of previously signed messages to distinguish genuine queries from forgeries. 
The real game does not; it verifies signatures as per the signature scheme’s definition. 
If a signature passes verification $(b= true)$, the ideal game checks 
whether it came from the signing oracle $(m \in S)$. 
If yes, the adversary has not forged anything new. 
If not, it forms a successful forgery attempt. 
This game-pair structure bridges the conceptual gap between strong unforgeability and indistinguishability, 
supporting robust reasoning for digital signatures used in protocols like \ra. 
This technique strengthens the security guarantees of \ra from correctness toward robust privacy. 
This shift allows us to move beyond isolated definitions of unforgeability 
toward composable, black-box security notions essential for verifying \ra protocols. 
\paragraph{Protocol-level reasoning.}
\vspace{-0.9em}
\begin{wrapfigure}{r}{0.30\textwidth}
  \centering
  \begin{tikzpicture}[
    scale=0.75, transform shape
  ]
	\node[] (real) at (0,0)  {
        $\mathcal{P}_{\phantom{\text{\ideal}}}^{\phantom{\Sigma}}$
	};
	\node[] (code-r) [below = 0cm of real] {
		\begin{minipage}{0.28\columnwidth}
			\begin{minted}[fontsize=\scriptsize,escapeinside=@@,autogobble]{coq}
@$\phantom{X}$@
@$\phantom{X}$@
prot(@$\msg$@)
  @$pk$@ @$\leftarrow$@ get_pk()
  @$\sig$@ @$\leftarrow$@ sign(@$\msg$@)
  @$b$@ @$\leftarrow$@
  ver_sig(@$pk,\sig,\msg$@)
  return (@$pk,\sig,b$@)
@$\phantom{X}$@
@$\phantom{X}$@
@$\phantom{X}$@
           \end{minted}
		\end{minipage}
	};

	\draw[] 
    (code-r.north west) |- 
    (real.north) -| 
    (code-r.north east);
	\draw[] 
    (code-r.north west) |- 
    (real.south) -| 
    (code-r.north east);
	\draw[] 
    (code-r.north east) |- 
    (code-r.south) -| 
    (code-r.north west);

    \draw[draw=lBlue, thick,<-] 
    ([xshift=0.05cm,yshift=-0.7cm]real.east) -- 
    node[above]{$m$} 
    ([xshift=1.05cm,yshift=-0.7cm]real.east);

    \draw[draw=lBlue, thick, ->] 
    ([xshift=0.05cm,yshift=-3.2cm]real.east) -- 
    node[above]{($pk$, $\sig$, $b$)}   
    ([xshift=1.05cm,yshift=-3.2cm]real.east);

    \node[draw] (comp1) [below left = 3.5cm and -0.3cm  of real] {
        $\mathcal{P} \circ \mathcal{G}_{\text{\real}}^{\Sigma}$
    };

    \node[draw] (comp2) [below right = 3.5cm and -0.3cm of real] {
       $\mathcal{P} \circ \mathcal{G}_{\text{\ideal}}^{\Sigma}$
    };

	\node[] (adv) [below = 5.3cm of real]  {
		$\mathcal{A}$ 
	};

	\draw[->] 
    (adv.north) -- 
    ([yshift=0.2cm]adv.north) -| 
    ([xshift=0.3cm,yshift=-0.1cm]comp1.south);
	
    \draw[->] 
    (adv.north) -- 
    ([yshift=0.2cm]adv.north) -| 
    ([xshift=-0.3cm,yshift=-0.1cm]comp2.south);

	\node[] (win) [above = 0.4cm of adv]  {
		 real\hspace{0.8cm}🏆\hspace{0.7cm}ideal
	};
\end{tikzpicture}
	\caption{Protocol}
  \label{fig:protocol}
\end{wrapfigure}
\vspace{0.9em}
Our formalization uses \ssprove to verify individual primitives and 
lift these guarantees to the protocol level.
The protocol-level abstraction
(Figure~\ref{fig:protocol}) treats signatures as black-box functionalities, 
enabling secure integration into broader systems such as \ra. 
Rather than verifying only functional correctness, this abstraction supports reasoning about composable security, 
capturing indistinguishability even when the protocol comprises other components.
The protocol calls \egetpk, \esign, and \eversig as modular components within a single routine, 
reflecting a structured composition that supports formal, game-based reasoning in \ssprove. 
This technique shows that replacing the signature module with an indistinguishable 
ideal version leaves the entire protocol indistinguishable from an external adversary.
%
%
Our formal development revealed that even properties like collision 
resistance at the primitive level become necessary for overall \ra security, though not required for correctness.
While \ssprove provides essential structure, bridging hand-written cryptographic arguments and 
machine-verified proofs remains challenging—especially when capturing game transformations and reductions. 
Our work addresses this by reconciling indistinguishability-based reasoning 
with formal proofs that scale to realistic protocol security.

Having defined the theoretical foundations of digital signature security, 
we now present their specification and verification. 
Since \ra builds upon digital signatures, we first formally specify digital signatures. 
Using the \ssprove framework, we modularize key generation, 
signing, and verification, proving the indistinguishability 
of \emph{real} and \emph{ideal} signature protocols. 
This modular development lays the groundwork for the 
formal security analysis of \ra in the next section.

Our formal development of digital signatures is organized 
around two components: the modular composition of protocol packages and 
the formal proof of their security properties. 
Each component is modeled as an \ssprove package to support modular and abstract reasoning. 
Specifically:
\vspace{-0.5em}
\begin{itemize}
  \item The \pkeygen package handles key generation.
  \item The \psigprim package implements the signing and verification primitives.
  \item The \psigprot package composes these primitives into a complete signature protocol.
\end{itemize}
\vspace{-0.5em}
The design centers around the \pkeygen package, 
abstracting the key generation process while enabling the 
specification to be later instantiated with 
concrete cryptographic algorithms such as RSA or ECDSA. 
Appendix~\ref{sec:rsa} provides an example of instantiation for RSA signatures.

Figure~\ref{fig:overview} illustrates the modular composition. 
Each box represents an \ssprove package, with arrows indicating 
the composition flow: the \pkeygen package generates keys, 
which \psigprim takes to perform signing and verification operations. 
\psigprot orchestrates these components to implement the full signature protocol.
This modular structure suffices to establish strong security guarantees. 
We formally prove that sEUF-CMA implies indistinguishability 
between the real and ideal signature protocols. 
Our development reduces the security of remote attestations to the security of secure signatures. 
The core intuition behind the proof is rooted in the sEUF-CMA property. 
Suppose an adversary cannot forge a signature on any new message. 
In that case, all accepted signatures must have been generated 
via the signing interface, rendering the behavior of the real and 
ideal protocols indistinguishable from the adversary’s point of view.
Conversely, if an adversary could distinguish between the two protocols, 
it must have found a valid signature on a message not generated by the signing interface, 
violating strong unforgeability. 
Thus, indistinguishability and strong unforgeability are tightly linked.

We formalize this reasoning and prove the following results. 
Theorem~\ref{theo:indist-signature} establishes Perfect Indistinguishability between the real and ideal digital signature protocols. 
Theorem~\ref{theo:redprot} establishes a reduction theorem showing that the security of remote attestation is at least as strong as the security of the underlying digital signatures. 
Theorem~\ref{theo:indist-ra} combines the two results to conclude that remote attestation satisfies strong existential unforgeability.
\vspace{-0.7em}
\begin{figure}
	\centering
	\scalebox{0.68}{
\begin{tikzpicture}[
    pkg/.style={draw,align=center,minimum height=1.5cm, minimum width=1.5cm},
    prot/.style={draw,align=center,minimum height=1.1cm},
    real/.style={fill=lGray!20},
    ideal/.style={fill=lYellow!20},
    imp/.style={rounded corners,draw,align=center,minimum height=0.5cm, minimum width = 3cm},
    imp2/.style={rounded corners,draw,align=center,minimum height=0.5cm, minimum width = 11cm},
    node distance=0.3cm,
    font=\scriptsize
  ]
  \node[pkg] (kg1) at (0,0) { \texttt{KeyGen}\\(Fig~\ref{fig:keygen:package}) };

  \node[pkg,dashed] (rsa) [below right = 1.5cm and -1cm  of kg1] { $\Sigma$ := RSA\\(\ref{sec:rsa}) };
  
    \node[pkg,real]  (sig1) [above right = 0cm and 1.75cm of kg1] { 
        \texttt{SigPrim}\\
        (Game~\ref{game:seuf-cma},Fig~\ref{fig:sigprim:packages})
    };
    \node[pkg,ideal] (sig2) [below right = 0cm and 1.75cm of kg1] {
        \texttt{SigPrim}\\
        (Game~\ref{game:seuf-cma},Fig~\ref{fig:sigprim:packages}) 
    };
 
    \node[pkg,real]  (sigprot1) [right = 1.5cm of sig1] { 
        \sigprot\\
        (Fig~\ref{fig:protocol})
    };
    \node[pkg,ideal] (sigprot2) [right = 1.5cm of sig2] { 
        \sigprot\\
        (Fig~\ref{fig:protocol}) 
    };
 
    \node[pkg,real]  (attprot3) [right = 1cm of sigprot1] { 
        \attprot\\
        (Fig~\ref{fig:att:prot})
    };
    \node[pkg,ideal] (attprot4) [right = 1cm of sigprot2] { 
        \attprot\\
        (Fig~\ref{fig:att:prot})
    };
 
    \node[align=center] (ind1) [right = 0.25cm of attprot3] { $\pindist$\\(Lem~\ref{lem:real}) };
    \node[align=center] (ind2) [right = 0.25cm of attprot4] { $\pindist$\\(Lem~\ref{lem:ideal}) };
    
    \node[pkg,real]  (attprot1) [right = 0.25cm of ind1] { 
        \attprot$^\prime$\\
        (Fig~\ref{fig:att:protprim})
    };
    \node[pkg,ideal] (attprot2) [right = 0.25cm of ind2] { 
        \attprot$^\prime$\\
        (Fig~\ref{fig:att:protprim})
    };
 
    \node[pkg,real]  (ra1) [right =1.5cm of attprot1] { 
        \texttt{AttPrim}\\
        (Fig~\ref{fig:att:prim}) 
    };
    \node[pkg,ideal] (ra2) [right = 1.5cm of attprot2] {  
        \texttt{AttPrim}\\
        (Fig~\ref{fig:att:prim}) 
    };

    \draw[]
    ([xshift=0.05cm]kg1.east)
    --node[midway,above]{\texttt{key\_gen}}
    ([xshift=1.25cm]kg1.east);
  
    \draw[->] ([xshift=1.25cm]kg1.east) |- ([xshift=-0.05cm]sig1.west);
    \draw[->] ([xshift=1.25cm]kg1.east) |- ([xshift=-0.05cm]sig2.west);
  
    \draw[->]
    ([xshift=0.05cm,yshift=0.5cm]sig1.east)
    -- node[midway,above]{\texttt{get\_pk}}
    ([xshift=-0.05cm,yshift=0.5cm]sigprot1.west);
    \draw[->]
    ([xshift=0.05cm,yshift=0.5cm]sig2.east)
    -- node[midway,above]{\texttt{get\_pk}}
    ([xshift=-0.05cm,yshift=0.5cm]sigprot2.west);

    \draw[->]
    ([xshift=0.05cm,yshift=0cm]sig1.east)
    --node[midway,above]{\texttt{sign}}
    ([xshift=-0.05cm,yshift=0cm]sigprot1.west);
    \draw[->]
    ([xshift=0.05cm,yshift=0cm]sig2.east)
    --node[midway,above]{\texttt{sign}}
    ([xshift=-0.05cm,yshift=0cm]sigprot2.west);
  
    \draw[->]
    ([xshift=0.05cm,yshift=-0.5cm]sig1.east)
    --node[midway,above]{\texttt{ver\_sig}}
    ([xshift=-0.05cm,yshift=-0.5cm]sigprot1.west);
    \draw[->]
    ([xshift=0.05cm,yshift=-0.5cm]sig2.east)
    --node[midway,above]{\texttt{ver\_sig}}
    ([xshift=-0.05cm,yshift=-0.5cm]sigprot2.west);

    \draw[<-]     
    ([xshift=0.05cm,yshift=0.5cm]attprot1.east)
    -- node[midway,above]{\texttt{get\_pk}} ([xshift=-0.05cm,yshift=0.5cm]ra1.west);
    \draw[<-]     
    ([xshift=0.05cm,yshift=0.5cm]attprot2.east)
    -- node[midway,above]{\texttt{get\_pk}} ([xshift=-0.05cm,yshift=0.5cm]ra2.west);
 
    \draw[<-]     
    ([xshift=0.05cm,yshift=0cm]attprot1.east)
    --node[midway,above]{\texttt{attest}} ([xshift=-0.05cm,yshift=0cm]ra1.west);
    \draw[<-]     
    ([xshift=0.05cm,yshift=0cm]attprot2.east)
    --node[midway,above]{\texttt{attest}} ([xshift=-0.05cm,yshift=0cm]ra2.west);
   
    \draw[<-]
    ([xshift=0.05cm,yshift=-0.5cm]attprot1.east)
    --node[midway,above]{\texttt{ver\_att}} ([xshift=-0.05cm,yshift=-0.5cm]ra1.west);
    \draw[<-]
    ([xshift=0.05cm,yshift=-0.5cm]attprot2.east)
    --node[midway,above]{\texttt{ver\_att}} ([xshift=-0.05cm,yshift=-0.5cm]ra2.west);

    \draw[->]
    ([xshift=-0.5cm,yshift=0.05cm]sig1.north) -- 
    ([xshift=-0.5cm,yshift=1.5cm]sig1.north)  -- 
    node[midway,above] {\texttt{get\_pk}}
    ([xshift=1.5cm,yshift=1.5cm]ra1.north east) --  
    ([xshift=1.5cm,yshift=-0.5cm]ra1.east) |- 
    ([xshift=0.05cm,yshift=-0.5cm]ra1.east);
    \draw[->]
    ([xshift=0cm,yshift=0.05cm]sig1.north) -- 
    ([xshift=0cm,yshift=1cm]sig1.north) -- 
    node[midway,above] {\texttt{sign}}
    ([xshift=1cm,yshift=1cm]ra1.north east) --  
    ([xshift=1cm,yshift=0cm]ra1.east) |- 
    ([xshift=0.05cm,yshift=0cm]ra1.east);
    \draw[->]
    ([xshift=0.5cm,yshift=0.05cm]sig1.north) -- 
    ([xshift=0.5cm,yshift=0.5cm]sig1.north) -- 
    node[midway,above] {\texttt{ver\_sig}}
    ([xshift=0.5cm,yshift=0.5cm]ra1.north east) --  
    ([xshift=0.5cm,yshift=0.5cm]ra1.east) |- 
    ([xshift=0.05cm,yshift=0.5cm]ra1.east);
    \draw[->]
    ([xshift=-0.5cm,yshift=-0.05cm]sig2.south) -- 
    ([xshift=-0.5cm,yshift=-1.5cm]sig2.south)  -- 
    node[midway,below] {\texttt{get\_pk}}
    ([xshift=1.5cm,yshift=-1.5cm]ra2.south east) --  
    ([xshift=1.5cm,yshift=0.5cm]ra2.east) |- 
    ([xshift=0.05cm,yshift=0.5cm]ra2.east);
    \draw[->]
    ([xshift=0cm,yshift=-0.05cm]sig2.south) -- 
    ([xshift=0cm,yshift=-1cm]sig2.south) -- 
    node[midway,below] {\texttt{sign}}
    ([xshift=1cm,yshift=-1cm]ra2.south east) --  
    ([xshift=1cm,yshift=0cm]ra2.east) |- 
    ([xshift=0.05cm,yshift=0cm]ra2.east);
    \draw[->]
    ([xshift=0.5cm,yshift=-0.05cm]sig2.south) -- 
    ([xshift=0.5cm,yshift=-0.5cm]sig2.south) -- 
    node[midway,below] {\texttt{ver\_sig}}
    ([xshift=0.5cm,yshift=-0.5cm]ra2.south east) --  
    ([xshift=0.5cm,yshift=-0.5cm]ra2.east) |- 
    ([xshift=0.05cm,yshift=-0.5cm]ra2.east);

    \draw[->] 
    ([xshift=0.05cm]sigprot1.east) --
    node[midway,above] { \texttt{prot} }
    ([xshift=-0.05cm]attprot3.west);
    \draw[->] 
    ([xshift=0.05cm]sigprot2.east) --
    node[midway,above] { \texttt{prot} }
    ([xshift=-0.05cm]attprot4.west);

    \path[dashed,white]
    (sigprot1.south)
    edge node[midway,align=center,text=black] (spindist) {
        \texttt{\pindist}\\
        (Fig~~\ref{fig:protocol},Thm~\ref{theo:indist-signature})
    }
    (sigprot2.north);

    \path[dashed,white]
    (attprot3.south)
    edge node[midway,left,align=center,text=black] (rared) {
        \textcolor{SomeGreen}{$\leq$}\\(Thm\ref{theo:redprot})
    }
    node[midway,right,align=center,text=black] (rapindist) {
        \texttt{\pindist}\\(Thm\ref{theo:indist-ra})
    }
    (attprot4.north);

    \begin{scope}[on background layer]
        \draw[dashed,SomeGreen] (spindist.east) -- (ind1.west);
        \draw[dashed,SomeGreen] (spindist.east) -- (ind2.west);
        \draw[dashed,SomeGreen] (ind1.south) -- (ind2.north);
    \end{scope}

    \node[fill=white] at (kg1.north west) { $\Sigma$ };
    \node[fill=white] at (sig1.north west) { $\Sigma$ };
    \node[fill=white] at (sig2.north west) { $\Sigma$ };

  \end{tikzpicture}
  
  }
	\caption{
  Modular specification of \ra using \ssprove package composition. 
  Real-world protocols are depicted in grey, while idealized protocols are shown in yellow. 
  The specification builds upon the modular structure of 
  digital signatures to establish formal security guarantees 
  through indistinguishability proofs.}\label{fig:overview}
\end{figure}
\vspace{-0.7em}
\subsection{Key Generation}
Following the definition of signatures, we formalize key generation.
%
The \pkeygen package,
abstracts the key generation process over arbitrary finite 
types for the secret key (\lseckey) and public key (\lpubkey). 
The package is parameterized by a key generation algorithm $\Sigma$.\pkeygen, 
which samples key pairs from a distribution.
%
%
%
\begin{wrapfigure}{r}{0.31\textwidth}
  \centering
  \begin{tikzpicture}[
  scale=0.75,
  transform shape
  ]
\node[] (P) at (0,0)  {
  \pkeygen
};

\node[] (c) [below = 0.5cm of P] {
\begin{minipage}{0.3\columnwidth}
\begin{minted}[fontsize=\scriptsize,escapeinside=&&,autogobble]{coq}
key_gen ():

  (sk,pk) <- &$\Sigma$&.KeyGen tt ;;
  #put @&$\mathcal{SK}$& sk ;;
  #put @&$\mathcal{PK}$& pk ;;

  #ret (sk,pk)
\end{minted}
\end{minipage}
};

\node[] (s) [above right = 0cm and 0cm of c.north west] {
  \begin{minipage}{0.35\columnwidth}
    \begin{minted}[fontsize=\scriptsize,escapeinside=&&,autogobble]{coq}
      &$\mathcal{SK}$& : SecKey, &$\mathcal{PK}$& : PubKey
    \end{minted}
  \end{minipage}
};

    \draw[->]
    ([xshift=0.2cm,yshift=-0cm]c.east) --
    ([xshift=1.8cm,yshift=-0cm]c.east)
        node[midway,above]{ \mintinline{Coq}{key_gen} };

    \draw[] (c.north west) -- (c.north east);
    \draw[] (c.north west) |- (s.north) -| (c.north east);
    \draw[] (s.north west) |- (P.north) -| (c.north east);
    \draw[] (c.north east) |- (c.south) -| (c.north west);

\end{tikzpicture}
  \caption{A \pkeygen \\ Package }
  \label{fig:keygen:package}
\end{wrapfigure}
%
%
Our formalization models key generation as an explicit sampling procedure, 
which aligns with the sEUF-CMA experiment and simplifies the reduction proofs in SSProve, 
where key generation produces probabilistically independent keys%
\footnote{
  Practical schemes adopt deterministic key derivation for efficiency or side-channel resistance. 
  However, our focus is foundational, and our formalization concentrates only on 
  proving cryptographic soundness right now. 
}.
This technique ensures the adversary cannot exploit correlations 
when attempting to forge signatures.
This design choice also ensures compatibility with the security game framework, 
where packages must use explicit probabilistic operations and cannot rely on external sources of randomness.
Therefore, the \pkeygen package exports the procedure 
\ekgen, 
which samples a key pair using $\Sigma$.\pkeygen, stores it in the local state, and returns the result. 
The signing and verification procedures later retrieve this stored key material. 
Crucially, $\Sigma$.\pkeygen must be polymorphic in this state, ensuring that the sampling 
process is side-effect-free for the external state of the composed packages. 
This technique enables the clean composition of \pkeygen with the signature primitives while preserving soundness guarantees.

Having defined key generation, we now formalize the signing and verification components.
%
%
%
\subsection{Signature Primitives}
We now define the signing and verification primitives abstractly over Message and Signature types. 
The signing function $\Sigma$\icoq{.Sign} takes a secret key and a message and outputs a signature. 
The verification function $\Sigma$.\icoq{VerSig} takes a public key, a message, 
and a signature and outputs a Boolean indicating whether the signature is valid. 
Both operations are deterministic and do not modify the package state.

The following hypothesis captures the functional correctness of the signature scheme:
\begin{hypothesis}[functional correctness]\label{hypo:funCorssprove}
\[   \forall m\, sk\, pk,
  ((sk, pk) \leftarrow \Sigma.\text{\pkeygen}) \rightarrow \\
  \Sigma.\text{\icoq{VerSig}}\, pk (\Sigma\text{\icoq{.Sign}}\, sk\, m)\, m == true.
\]
\end{hypothesis}
\vspace{-1.5em}
\begin{figure}
  \centering
  \begin{tikzpicture}[
	scale=0.75, transform shape
	]
	\node[] (real) at (0,0)  {
		\psigprim$_{\text{\real}}$
	};
	\node[] (code-r) [below = 0.75cm of P] {
		\begin{minipage}{0.29\columnwidth}
			\begin{minted}[fontsize=\scriptsize,escapeinside=&&,autogobble]{coq}
get_pk() :=
  #import key_gen ;;

  (_,pk) <- key_gen tt ;;
  #ret pk

sign(m) :=
  sk <- #get @&$\mathcal{SK}$& ;;
  let &$\sigma$& := &$\Sigma$&.Sign sk m ;;



  #ret &$\sigma$&

ver_sig(m,s) :=
  pk <- #get @&$\mathcal{PK}$& ;;
  #ret &$\Sigma$&.VerSig pk s m
			\end{minted}
		\end{minipage}
	};
	\node[] (state-r) [above right = 0cm and 0cm of code-r.north west] {
	\begin{minipage}{0.29\columnwidth}
		\begin{minted}[fontsize=\scriptsize,escapeinside=&&,autogobble]{coq}
&$\mathcal{SK}$& : SecKey, &$\mathcal{PK}$& : PubKey
&\phantom{$\mathcal{SIG}$}&
		\end{minted}
	\end{minipage}
	};

  \draw[] (code-r.north west) -- (code-r.north east);
	\draw[] (code-r.north west) |- (state-r.north) -| (code-r.north east);
	\draw[] (state-r.north west) |- (real.north) -| (code-r.north east);
	\draw[] (code-r.north east) |- (code-r.south) -| (code-r.north west);


    \node[] (ideal) [right = 2.1 cm of real]  {
      \psigprim$_{\text{\ideal}}$
    };
    \node[] (code-i) [below = 0.75cm of ideal] {
    	\begin{minipage}{0.32\columnwidth}
    		\begin{minted}[fontsize=\scriptsize,escapeinside=&&,autogobble,highlightlines={10,11,12,16,17},highlightcolor=lYellow!20]{coq}
get_pk() :=
  #import key_gen ;;

  (_,pk) <- key_gen tt ;;
  #ret pk

sign(m) :=
  sk <- #get @&$\mathcal{SK}$& ;;
  let s := &$\Sigma$&.Sign sk m ;;
  S <- #get @&$\mathcal{SIG}$& ;;
  let S' := S &$\cup$& {(m,s)} ;;
  #put @&$\mathcal{SIG}$& S' ;;
  #ret s

ver_sig(m, s) :=
  S <- #get @&$\mathcal{SIG}$& ;;
  #ret ( (m,s) &$\in$& S )
    		\end{minted}
    	\end{minipage}
    };
    \node[] (state-i) [above right = 0cm and 0cm of code-i.north west] {
    	\begin{minipage}{0.32\columnwidth}
    		\begin{minted}[fontsize=\scriptsize,escapeinside=&&,autogobble,highlightlines={2},highlightcolor=lYellow!20]{coq}
&$\mathcal{SK}$&: SecKey, &$\mathcal{PK}$&: PubKey,
&$\mathcal{SIG}$&: (S(Message x Signature))
    		\end{minted}
    	\end{minipage}
    };

    \draw[] (code-i.north west) -- (code-i.north east);
    \draw[] (code-i.north west) |- (state-i.north) -| (code-i.north east);
    \draw[] (state-i.north west) |- (ideal.north) -| (code-i.north east);
    \draw[] (code-i.north east) |- (code-i.south) -| (code-i.north west);

\end{tikzpicture}
  \caption{The \real and \ideal Signature Primitives packages.}
  \label{fig:sigprim:packages}
\end{figure}
\vspace{-1.0em}
That is, for any key pair generated 
by $\Sigma$.\pkeygen and any message $m$,
signing, and then verifying should succeed.

We implement two versions of the signature primitive packages (Figure~\ref{fig:sigprim:packages}). 
The real version directly implements the signing and verification operations, 
invoking $\Sigma$\icoq{.Sign} and $\Sigma$.\icoq{VerSig} respectively. 
The ideal version tracks an internal set \icoq{S} of signed messages and 
accepts a signature during verification only if it 
corresponds to a message previously signed through the signing interface.
Both packages rely on the \egetpk procedure from \pkeygen to retrieve the \icoq{pk}. 
With the primitives in place, we now describe the composition of the real and ideal protocols 
and the security proof that establishes their indistinguishability.
\subsection{Indistinguishability Proof}
We construct the real and ideal digital signature protocols to formalize the 
security property by composing the primitives with the key generation package. 
The composed protocols are defined as follows:
\vspace{-0.9em}
\begin{definition}[\psigprot$_{\text{\real}}$]\label{def:sigprotR}
  \( \{\)package \psigprot $\circledcirc$ \, \psigprim$_{\text{\real}}$ $\circ$ \, \pkeygen\( \}\).
\end{definition}
\begin{definition}[\psigprot$_{\text{\ideal}}$]\label{def:sigprotI} 
  \( \{\)package \psigprot $\circledcirc$ \, \psigprim$_{\text{\ideal}}$ $\circ$ \, \pkeygen\( \}\).
\end{definition}
\vspace{-0.5em}
The wrapper package \psigprot relabels the procedures of the 
signature primitives to provide a unified interface (special notation $\circledcirc$). 
Specifically, it renames the \esign operation to \icoq{challenge} and 
the verification operation \eversig to \icoq{verify}%
\footnote{
In \ssprove, this is not necessary because procedure names map
to a particular natural number.
Mapping different procedure names to the same natural numbers
establishes the link without an auxiliary package for renaming
procedures.
}. 
This relabeling ensures that the composed packages match the 
interface and are comparable to the security proof.
We then establish our main result:
\vspace{-0.2em}
\begin{theorem}[Perfect Indistinguishability of Digital Signatures]\label{theo:indist-signature}
  For every adversary \A, the advantage to distinguish the packages
  \psigprot$_{\text{\real}}$ and \psigprot$_{\text{\ideal}}$ is $0$:

  \qquad \qquad \qquad \qquad \psigprot$_{\text{\real}}$ \pindist \psigprot$_{\text{\ideal}}$.
\end{theorem} 
\vspace{-0.2em}
\vspace{-0.5em}
\begin{proof}
  The proof proceeds by establishing an invariant \icoq{heap_ignore} (Figure~\ref{fig:ind:proof})
  that the heap states of the real and ideal packages are identical 
  up to an irrelevant location $\mathcal{SIG}$.
  As the protocols evolve, this invariant is preserved through every execution step, ensuring that signing 
  and verification operations behave identically from the adversary’s perspective.
  The critical point in the proof is to show that any observable difference 
  between the real and ideal games would require a signature to verify 
  correctly in the real protocol without having been produced by 
  the signing interface in the ideal protocol. 
  However, this is ruled out by the signature scheme's functional correctness property 
  (Definition\ref{def:sig:correct}). 
  Note that the real game accepts any syntactically valid signature, 
  and the ideal game only accepts oracle-issued ones. 
  This semantic difference does not make much difference in our formalization 
  because under the sEUF-CMA, the adversary cannot produce a new valid signature on any message, 
  so any accepted signature must have originated from the oracle. 
  Therefore, the adversary gains no distinguishing power. 
  %
  %
  %
  so any accepted signature must have originated from the oracle. 
  %
  However, \ra formalization (Section~\ref{sec:vra}) assumes injectivity 
  of the hash function $\mathcal{H}$ and
  ensure that each challenge-state pair maps to a unique message. 
  This prevents adversaries from reusing signatures across different attestation sessions. 
  The combination of injectivity and unforgeability guarantees that an adversary cannot 
  substitute or reapply old signatures for new challenges or states. 
\end{proof}
\vspace{-0.3em}
%
%
\vspace{-0.9em}
\begin{figure}
  \begin{minipage}{0.25\columnwidth}
  \begin{minted}[escapeinside=@@,fontsize=\scriptsize,autogobble]{Coq}
  @$\vdash$@
  { @$\lambda$@ (s@$_1$@, s@$_2$@),
    let inv := heap_ignore @$\{\mathcal{SIG}\}$@ in
    let h   := inv @$\rtimes$@ rm_rhs @$\mathcal{SIG}$@ S in
    let h'  := set_rhs @$\mathcal{SIG}$@ (S @$\cup$@ @$\{$@(@$\Sigma$@.Sign sk m, m)@$\}$@ h in
    let h'' := h' @$\rtimes$@ rm_rhs @$\mathcal{SIG}$@ S' in
    h'' (s@$_1$@, s@$_2$@)) }
  ret (pk, @$\Sigma$@.Sign sk m, @$\Sigma$@.VerSig pk (@$\Sigma$@.Sign sk m) m)
  @$\approx$@
  ret (pk, @$\Sigma$@.Sign sk m, (@$\Sigma$@.Sign sk m, m) @$\in$@ S')
  { @$\lambda$@ (s@$_1$@,r@$_1$@)
      (s@$_2$@,r@$_2$@), r@$_1$@ = r@$_2$@ @$\land$@ heap_ignore @$\{\mathcal{SIG}\}$@ (s@$_1$@, s@$_2$@) }
  \end{minted}
  \end{minipage}
  \caption{
    %
    An invariant \icoq{heap_ignore}.
  }\label{fig:ind:proof}
  \end{figure}
  \vspace{-0.8em}
\section{Formal development of Remote Attestation (\ra)}\label{sec:vra}
\vspace{-0.5em}
This section formalizes \ra and proves its security against forgery. 
We define packages, describe the protocol, and demonstrate that \ra inherits security 
from digital signatures through a sequence of reductions. 
By reducing its security to that of the underlying digital signature scheme, 
we aim to establish perfect indistinguishability for \ra.
\vspace{-0.6em}
\subsection{Packages}
We model the \ra setup using \ssprove packages that abstract over cryptographic primitives 
and their interactions with an adversary. 
\vspace{-1.2em}
\begin{figure}\centering
  \begin{subfigure}[b]{0.44\textwidth}
      \centering
    \resizebox{1\linewidth}{!}{\begin{tikzpicture}[
	scale=0.75, transform shape
	]
	\node[] (P) at (0,0)  {
		 AttProt
	};
	\node[] (c) [below = 0.5cm of P] {
		\begin{minipage}{0.50\columnwidth}
            \begin{minted}[fontsize=\scriptsize,escapeinside=&&,highlightlines={4,5},highlightcolor=lYellow!20]{coq}
prot(c)
  #import sig_prot;;

  s <- #get @&$\mathcal{ST}$&
  #let m := &$\mathcal{H}$& s c ;;
  a := sig_prot m ;;
  
  #ret a
			\end{minted}
		\end{minipage}
	};
	\node[] (s) [above right = 0cm and 0cm of c.north west] {
		\begin{minipage}{0.50\columnwidth}
            \begin{minted}[fontsize=\scriptsize,escapeinside=&&,autogobble]{coq}
	&$\mathcal{ST}$&: State
			\end{minted}
		\end{minipage}
	};

	\draw[->] ([xshift=-1.8cm]c.west) -- ([xshift=-0.2cm]c.west)
	 node[midway,above] { \mintinline{Coq}{sig_prot} };
	
    \draw[->] ([xshift=0.2cm]c.east) -- ([xshift=1.8cm]c.east) 
    node[midway,above] { \mintinline{Coq}{prot} };

	\draw[] (c.north west) -- (c.north east);
	\draw[] (c.north west) |- (s.north) -| (c.north east);
	\draw[] (s.north west) |- (P.north) -| (c.north east);
	\draw[] (c.north east) |- (c.south) -| (c.north west);
	
\end{tikzpicture}}
	\caption{Using the signature \\protocol}
	\label{fig:att:prot}
  \end{subfigure}
  \hfill
  \begin{subfigure}[b]{0.44\textwidth}
      \centering
    \resizebox{1\linewidth}{!}{\begin{tikzpicture}[
	scale=0.75, transform shape
	]
	\node[] (P) at (0,0)  {
        Att\pprot{$^\prime$}
	};
	\node[] (c) [below = 0.5cm of P] {
		\begin{minipage}{0.50\columnwidth}
			\begin{minted}[fontsize=\scriptsize,escapeinside=&&,autogobble]{coq}
def prot (c):
  #import key_gen ;;
  #import attest ;;
  #import verify&$^a$& ;;

  pk <- get_pk tt ;;
  s <- attest c ;;
  b <- verify&$^a$& s c ;;

  #ret (pk,s,b)
			\end{minted}
		\end{minipage}
	};
	\node[] (s) [above right = 0cm and 0cm of c.north west] {
        \begin{minipage}{0.50\columnwidth}
            \begin{minted}[fontsize=\scriptsize,escapeinside=&&,autogobble]{coq}
				  &$\{ \}$&
			\end{minted}
		\end{minipage}
	};

	\draw[->] 
    ([xshift=-1.8cm,yshift=-0.7cm]c.north west) -- 
    ([xshift=-0.2cm,yshift=-0.7cm]c.north west)
	    node[midway,above] { \mintinline{Coq}{get_pk} };
	
	\draw[->] 
    ([xshift=-1.8cm,yshift=-0.2cm]c.west) -- 
    ([xshift=-0.2cm,yshift=-0.2cm]c.west)
        node[midway,above] { \mintinline{Coq}{attest} };

	\draw[->] 
    ([xshift=-1.8cm,yshift=0.2cm]c.south west) -- 
    ([xshift=-0.2cm,yshift=0.2cm]c.south west)
        node[midway,above]{ \mintinline{Coq}{verify}$^a$ };
	
    \draw[->] 
    ([xshift=0.2cm,yshift=-0.2cm]c.east) -- 
    ([xshift=1.8cm,yshift=-0.2cm]c.east)
        node[midway,above]{ \mintinline{Coq}{prot} };

	\draw[] (c.north west) -- (c.north east);
	\draw[] (c.north west) |- (s.north) -| (c.north east);
	\draw[] (s.north west) |- (P.north) -| (c.north east);
	\draw[] (c.north east) |- (c.south) -| (c.north west);

\end{tikzpicture}}
	\caption{Using the attestation \\primivites}
	\label{fig:att:protprim}
  \end{subfigure}
  \caption{Two variants to construct the \ra protocol.}
  \label{fig:att}
\end{figure}
\vspace{-0.6em}
Figure~\ref{fig:att} defines two packages: \pattprot{$^\prime$}, 
representing the \ra primitives, and \pattprot, 
the \ra protocol built atop these primitives. 
Both packages depend on an abstract \ststate type $\mathcal{ST}$, 
representing the platform being attested. 
Their interfaces expose three key operations: first, \egetpka for public key access; 
second, \eattest for issuing attestations over challenges $c$, and 
the last is the \everifya that a challenge $c$ was attested with $a$. 
In Figure~\ref{fig:att:prim}, the package \pattprim directly imports the digital signature 
primitives from \psigprim (Figure~\ref{fig:sigprim:packages}). 
These include key generation, signing, and verification 
and are parameterized by a cryptographic hash function $\mathcal{H}$ that maps a challenge 
and system state to a message to be signed. 
On the other hand, \pattprot extends the digital signature protocol 
(Definitions~\ref{def:sigprotR}, \ref{def:sigprotI}). 
The attestation is a tuple that contains the hashed state and the 
respective signature over this state. 
This design ensures that the attestation is bound to both the current state 
and a fresh verifier-supplied challenge, thereby preventing replay attacks.
The \everifya recomputes the hash internally from $c$ and 
the current state, assuming the state is fixed—an assumption 
valid in static \ra settings like a secure boot 
(the system's configuration does not change during the attestation).
While more dynamic \ra would require verifying against 
externally supplied hashes, this distinction is irrelevant for 
indistinguishability: both approaches yield equivalent results in our model. 
We adopt the constant-state version for simplicity and alignment with common \ra setups.
Security relies on the indistinguishability of the two packages. 
If an adversary could distinguish \pattprot from \pattprot{$^\prime$}, 
the underlying signature scheme would be broken. 
The reduction leverages collision resistance by assuming \emph{injective} (\emph{uncurry Hash}) 
in \ssprove, 
ensuring that the hash uniquely binds the challenge and 
state and preventing forgery via hash collisions.
\vspace{-0.6em}
\begin{figure}
  \centering
  \begin{tikzpicture}[
	scale=0.75, transform shape
	]
	\node[] (P) at (0,0)  {
		 AttPrim
	};
	\node[] (c) [below = 0.4cm of P] {
		\begin{minipage}{0.22\columnwidth}
            \begin{minted}[fontsize=\scriptsize,escapeinside=&&,autogobble,highlightlines={8,9,15,16},highlightcolor=lYellow!20]{coq}
get_pk&$^a$&()
  #import get_pk ;;
  pk <- get_pk ;;
  #ret pk
  
attest(c)
  #import sign;;
  s <- #get @&$\mathcal{ST}$&
  #let m := &$\mathcal{H}$& s c ;;
  a <- sign m ;;
  #ret (a,m)

verify&$^a$&(c,a)
  #import verify ;;
  s <- #get @&$\mathcal{ST}$& ;;
  #let m := &$\mathcal{H}$& s c ;;
  #ret verify a m
			\end{minted}
		\end{minipage}
	};
	\node[] (s) [above right = 0cm and 0cm of c.north west] {
		\begin{minipage}{0.22\columnwidth}
            \begin{minted}[fontsize=\scriptsize,escapeinside=&&,autogobble]{coq}
	&$\mathcal{ST}$&: State
			\end{minted}
		\end{minipage}
	};

	\draw[->] ([xshift=-1.8cm,yshift=-1cm]c.north west) -- ([xshift=-0.2cm,yshift=-1cm]c.north west)
	    node[midway,above] { \mintinline{Coq}{get_pk} };
	
	\draw[->] ([xshift=-1.8cm,yshift=3cm]c.south west) -- ([xshift=-0.2cm,yshift=3cm]c.south west)
        node[midway,above] { \mintinline{Coq}{sign} };

	\draw[->] ([xshift=-1.8cm,yshift=1cm]c.south west) -- ([xshift=-0.2cm,yshift=1cm]c.south west)
        node[midway,above]{ \mintinline{Coq}{verify} };
	
    \draw[->] ([xshift=0.2cm,yshift=-1cm]c.north east) -- ([xshift=1.8cm,yshift=-1cm]c.north east)
        node[midway,above]{ \mintinline{Coq}{get_pk}$^a$ };

    \draw[->] ([xshift=0.2cm,yshift=3cm]c.south east) -- ([xshift=1.8cm,yshift=3cm]c.south east)
        node[midway,above]{ \mintinline{Coq}{attest} };

    \draw[->] ([xshift=0.2cm,yshift=1cm]c.south east) -- ([xshift=1.8cm,yshift=1cm]c.south east)
        node[midway,above]{ \mintinline{Coq}{verify}$^a$ };
	
	\draw[] (c.north west) -- (c.north east);
	\draw[] (c.north west) |- (s.north) -| (c.north east);
	\draw[] (s.north west) |- (P.north) -| (c.north east);
	\draw[] (c.north east) |- (c.south) -| (c.north west);
	
\end{tikzpicture}	
  \caption{Package for \ra primitives.}
  \label{fig:att:prim}
\end{figure}
\vspace{-0.9em}
\vspace{-0.4em}
\subsection{Security Reduction and Indistinguishability (IND) for Protocols}
\vspace{-0.4em}
To formally reduce the security of \ra to that of digital signatures, 
we define both \real and \ideal versions of the attestation protocol 
by composing the corresponding digital signature packages. 
In the following definitions, we construct four key packages:
\vspace{-0.5em}
\begin{definition}[\pattprot{$_{\text{\real}}^{\text{Prim}}$}]\label{def:attPrimR}

  \( \{\)package $\text{\pattprot}^{\prime}$ $\circledcirc$ \, \text{\pattprim} \, $\circ$ \, \text{\psigprim}$_{\text{\real}}$ $\circ$ \, \text{\pkeygen}\( \}\).
\end{definition}
\begin{definition}[\pattprot{$_{\text{\ideal}}^{\text{Prim}}$}]\label{def:attPrimI}

  \( \{\)package $\text{\pattprot}^{\prime}$ $\circledcirc$ \, \text{\pattprim} \, $\circ$ \, \text{\psigprim}$_{\text{\ideal}}$ $\circ$ \, \text{\pkeygen}\( \}\).
\end{definition}
\begin{definition}[\pattprot{$_{\text{\real}}^{\text{Prot}}$}]\label{def:attProtR}
  \( \{\)package \text{\pattprot} \, $\circledcirc$ \, \text{\psigprot}$_{\text{\real}}$   \( \}\).
\end{definition}
\begin{definition}[\pattprot{$_{\text{\ideal}}^{\text{Prot}}$}]\label{def:attProtI}
  \( \{\)package \text{\pattprot} \, $\circledcirc$ \, \text{\psigprot}$_{\text{\ideal}}$   \( \}\).
\end{definition}
\vspace{-0.4em}
The first two represent the primitive-level abstraction of remote attestation, 
composed of \real or \ideal digital signature primitives. 
The latter two represent the protocol-level abstraction, 
containing the full digital signature protocol interface.
With these definitions, we establish the indistinguishability of the \real and \ideal \ra protocols, 
assuming that the underlying signature schemes are indistinguishable. 
The reduction proceeds through a sequence of games, 
progressively replacing real components with their ideal counterparts, 
preserving behavioral equivalence at each step.
We formalize this as a refinement between the \real and \ideal packages:
\vspace{-0.4em}
\begin{lemma}[Perfect IND of \real Attestation Protocols]\label{lem:real}
    For every adversary \A, the advantage of distinguishing
    the \real packages for \ra is $0$:
    \vspace{-0.4em}
    \begin{center}
        \begin{minipage}{0.5\columnwidth}
    \begin{minted}[fontsize=\footnotesize,escapeinside=&&,autogobble]{coq}
AttProt&$_{\text{\real}}^{\text{Prim}}$& &$\pindist$& AttProt&$_{\text{\real}}^{\text{Prot}}$&.
    \end{minted}
    \end{minipage}
    \end{center}
\end{lemma}
\vspace{-0.4em}
\vspace{-0.4em}
\begin{proof}
    First, we will inline the procedures for the \ra primitives in \pattprot{$_{\text{\real}}^{_\text{Prim}}$}.
    Afterward, we inline \psigprot in the \pattprot{$_{\text{\real}}^{_\text{Prot}}$}.
    Recomputing the hash in \everifya is canceled
    because \psigprot reuses the message,
    in this case, the hashed state, as input to \esign and \eversig.
    The rest of the proof is obtained by applying the rules in relational Hoare logic.
\end{proof}
\vspace{-0.4em}
\vspace{-0.4em}
\begin{lemma}[Perfect IND for \ideal Attestation Protocols]\label{lem:ideal}
    For every adversary \A, the advantage of distinguishing
    the \ideal packages for \ra is $0$: 
    \begin{center}
        \begin{minipage}{0.5\columnwidth}
    \begin{minted}[fontsize=\footnotesize,escapeinside=&&,autogobble]{coq}
AttProt&$_{\text{\ideal}}^{\text{Prim}}$& &$\pindist$& AttProt&$_{\text{\ideal}}^{\text{Prot}}$&.
    \end{minted}
    \end{minipage}
    \end{center}
\end{lemma}
\vspace{-0.4em}
\begin{proof}
The proof reasoning is analogous to
the proof of Lemma~\ref{lem:real}.
\end{proof}
\vspace{-0.4em}
%
%
\vspace{-0.4em}
\begin{theorem}[Security reduction for \ra]\label{theo:redprot}
  For every adversary \A, the advantage of distinguishing 
  the packages \pattprot{$_{\text{\real}}^{\text{Prim}}$} and 
  \pattprot{$_{\text{\ideal}}^{\text{Prim}}$}
  is less than or equal to the advantage of distinguishing the packages: 
  \psigprot{$_{\text{\real}}$} and \psigprot{$_{\text{\ideal}}$}
  \vspace{-0.4em}
  \begin{center}
      \begin{minipage}{0.8\columnwidth}
     \begin{minted}[fontsize=\footnotesize,escapeinside=&&,autogobble]{coq}
  &$\forall$& A,&\spacer{0.1cm}& Adv A             AttProt&$_{\text{\real}}^{\text{Prim}}$& AttProt&$_{\text{\ideal}}^{\text{Prim}}$& &$\leq$& &\spacer{0.1cm}&
       Adv (A &$\circ$& AttProt) SigProt&$_{\text{\ideal}}$& SigProt&$_{\text{\real}}$&.
    \end{minted}
     \end{minipage}
  \end{center}
\end{theorem}
\vspace{-0.4em}
\vspace{-0.4em}
\begin{proof}
    \ssprove allows us to use the following equalities:
    \begin{center}
    \begin{minipage}{0.9\columnwidth}
    \begin{minted}[fontsize=\footnotesize,escapeinside=&&,autogobble,linenos]{coq}
  Adv (A &$\circ$& AttProt) SigProt&$_{\text{\ideal}}$& SigProt&$_{\text{\real}}$& &\spacer{0.2cm}&
= Adv A (AttProt &$\circ$& SigProt&$_{\text{\ideal}}$& (AttProt &$\circ$& SigProt&$_{\text{\real}}$&)&\spacer{0.2cm}&
= Adv A AttProt&$_{\text{\real}}^{\text{Prot}}$& AttProt&$_{\text{\real}}^{\text{Prot}}$&
    \end{minted}
    \end{minipage}
    \end{center}
    The first equality (Line~2) is achieved by reducing the lemma in \ssprove
    (as shown in their Lemma~2.3~\cite{ssprove}).
    The second equality (Line~3) is, by definition, of the 
    packages for remote attestation (Definitions~\ref{def:attPrimR},\ref{def:attPrimI},\ref{def:attProtR},\ref{def:attProtI}).
    Our goal then becomes
    \begin{minted}[fontsize=\footnotesize,escapeinside=&&,autogobble]{coq}
  &$\forall$& A, Adv A AttProt&$_{\text{\real}}^{\text{Prim}}$& AttProt&$_{\text{\ideal}}^{\text{Prim}}$& &$\leq$& Adv A AttProt&$_{\text{\real}}^{\text{Prot}}$& AttProt&$_{\text{\ideal}}^{\text{Prot}}$&.
    \end{minted}
    We define a triangle inequality
    (as shown in their Lemma~2.2~\cite{ssprove}), and use transitivity of the 
    inequality ($\leq$) to obtain:
    \begin{minted}[fontsize=\footnotesize,escapeinside=&&,autogobble]{coq}
  Adv A AttProt&$_{\text{\real}}^{\text{Prim}}$& AttProt&$_{\text{\real}}^{\text{Prot}}$&&\spacer{0.2cm}& + Adv A AttProt&$_{\text{\real}}^{\text{Prot}}$& AttProt&$_{\text{\ideal}}^{\text{Prot}}$&&\spacer{0.2cm}& 
             + Adv A AttProt&$_{\text{\ideal}}^{\text{Prot}}$& AttProt&$_{\text{\ideal}}^{\text{Prim}}$&&\spacer{0.2cm}& &$\leq$& Adv A AttProt&$_{\text{\real}}^{\text{Prot}}$& AttProt&$_{\text{\ideal}}^{\text{Prot}}$&.
    \end{minted}
    By symmetry of games and Lemmas~\ref{lem:real} 
    and \ref{lem:ideal}, our goal reduces to:
    \begin{minted}[fontsize=\footnotesize,escapeinside=&&,autogobble]{coq}
  0 + Adv A AttProt&$_{\text{\real}}^{\text{Prot}}$& AttProt&$_{\text{\ideal}}^{\text{Prot}}$& + 0 &$\leq$& Adv A AttProt&$_{\text{\real}}^{\text{Prot}}$& AttProt&$_{\text{\ideal}}^{\text{Prot}}$&.
  \end{minted}
  Clearly, this holds by the
  left and right identity of addition and
  the definition of $\leq$ itself.
\end{proof}
Based on this result, we claim
perfect indistinguishability for the \ra protocol.
\begin{theorem}[Perfect IND of \ra]\label{theo:indist-ra}
  For every adversary \A, the advantage of distinguishing the packages
  \pattprot{$_{\text{\real}}^{\text{Prim}}$} and \pattprot{$_{\text{\ideal}}^{\text{Prim}}$ is $0$}:
     \begin{center}
        \begin{minipage}{0.5\columnwidth}
    \begin{minted}[fontsize=\footnotesize,escapeinside=&&,autogobble]{coq}
AttProt&$_{\text{\real}}^{\text{Prim}}$& &$\pindist$& AttProt&$_{\text{\ideal}}^{\text{Prim}}$&.
    \end{minted}
    \end{minipage}
    \end{center}
\end{theorem}
\vspace{-0.5em}
\begin{proof}
    To prove the theorem 
    \begin{minted}[fontsize=\footnotesize,escapeinside=&&,autogobble]{coq}
  Adv A AttProt&$_{\text{\real}}^{\text{Prim}}$& AttProt&$_{\text{\ideal}}^{\text{Prim}}$& &$\leq$& 0
    \end{minted}
    we apply our 
    Reduction Theorem~\ref{theo:redprot}.
    The goal changes into:
     \begin{minted}[fontsize=\footnotesize,escapeinside=&&,autogobble]{coq}
Adv (A &$\circ$& AttProt) SigProt&$_{\text{\ideal}}$& SigProt&$_{\text{\real}}$& &$\leq$& 0
    \end{minted}
    This holds by perfect indistinguishability
    of digital signatures (Theorem~\ref{theo:indist-signature}),
    even for an attacker \A{} that runs the
    attestation protocol \pattprot.
\end{proof}
\vspace{-0.7em}
\subsection{Security Reduction for Primitives}\label{sec:coll}
We refine our reduction by linking the security of \ra primitives to that of digital signature primitives. 
Figure~\ref{fig:att:prim:game} defines the semantics of \ra primitives using an auxiliary state $\mathcal{Z}$ 
that maps challenges $c$ to attestations $a$. 
\vspace{-0.5em}
\begin{definition}[\psigprimattR]\label{def:prim1}
  \( \{\)package \pattprim $\circ$  \psigprim$_{\text{\real}}$ $\circ$  \pkeygen\( \}\).
\end{definition}
\begin{definition}[\psigprimattId]\label{def:prim2} 
  \( \{\)package \pattprim $\circ$  \psigprimId $\circ$  \pkeygen\( \}\).
\end{definition}
\begin{definition}[\pattprimsig]\label{def:prim3}

  \( \qquad \qquad \quad \qquad \qquad \qquad \{\)package \pattprimR $\circ$ \, \psigprimR $\circ$ \, \pkeygen\( \}\).
\end{definition}
\begin{definition}[\pattprimsigId]\label{def:prim4}

  \( \qquad \qquad \quad \qquad \qquad \qquad \{\)package \pattprimId $\circ$ \, \psigprimId $\circ$ \, \pkeygen\( \}\).
\end{definition}
\vspace{-0.6em}
The \everifya procedure queries this mapping. 
This package, denoted \pattprim{$_{\real}$}, coincides with the earlier \pattprim from Figure~\ref{fig:att:prim}.
In Definitions~\ref{def:prim1},~\ref{def:prim2},~\ref{def:prim3},~\ref{def:prim4}, 
we lift digital signature primitives into the \ra setting 
via the \pattprim package and 
use this to define both real and ideal \ra primitive games.
\vspace{-0.5em}
\begin{figure}
  \centering
  \begin{tikzpicture}[
	scale=0.85, transform shape
	]
	\node[] (real) at (0,0)  {
		\pattprim$_{\text{\real}}$
	};
	\node[] (code-r) [below = 0.75cm of P] {
		\begin{minipage}{0.23\columnwidth}
			\begin{minted}[fontsize=\scriptsize,escapeinside=&&,autogobble]{coq}
get_pk&$^a$&()
  &\kimport& get_pk ;;
  pk <- get_pk ;;
  &\kret& pk
 
attest(c)
  &\kimport& sign;;
  s <- &\kget& @&$\mathcal{ST}$&
  #let m := &$\mathcal{H}$& s c ;;
  a <- sign m ;;



  &\kret& (a,m)

verify&$^a$&(c,a)
  &\kimport& verify ;;
  s <- &\kget& @&$\mathcal{ST}$& ;;
  #let m := &$\mathcal{H}$& s c ;;
  &\kret& verify a m
	        \end{minted}
		\end{minipage}
	};
	\node[] (state-r) [above right = 0cm and 0cm of code-r.north west] {
	\begin{minipage}{0.23\columnwidth}
		\begin{minted}[fontsize=\scriptsize,escapeinside=&&,autogobble]{coq}
&$\mathcal{ST}$& : State
&$\phantom{\mathcal{Z}}$&
	\end{minted}
	\end{minipage}
	};

  \draw[] (code-r.north west) -- (code-r.north east);
	\draw[] (code-r.north west) |- (state-r.north) -| (code-r.north east);
	\draw[] (state-r.north west) |- (real.north) -| (code-r.north east);
	\draw[] (code-r.north east) |- (code-r.south) -| (code-r.north west);


    \node[] (ideal) [right = 1.5 cm of real]  {
      \pattprim$_{\text{\ideal}}$
    };
    \node[] (code-i) [below = 0.86cm of ideal] {
    	\begin{minipage}{0.29\columnwidth}
    		\begin{minted}[fontsize=\scriptsize,escapeinside=&&,autogobble,highlightlines={11,12,13,18,20},highlightcolor=lYellow!20]{coq}
get_pk&$^a$&()
  &\kimport& get_pk ;;
  pk <- get_pk ;;
  &\kret& pk
 
attest(c)
  &\kimport& sign ;;
  s <- &\kget& @&$\mathcal{ST}$& ;;
  #let m := &$\mathcal{H}$& s c ;;
  a <- sign m ;;
  Z <- &\kget& @&$\mathcal{Z}$& ;;
  let Z' := Z &$\cup$& {(c,a)} ;;
  &\kput& @&$\mathcal{Z}$& Z' ;;
  &\kret& (a,m)

verify&$^a$&(c,a)
  
  Z <- &\kget& @&$\mathcal{Z}$& ;;

  &\kret& ( (c,a) &$\in$& Z )
    		\end{minted}
    	\end{minipage}
    };
    \node[] (state-i) [above right = 0cm and 0cm of code-i.north west] {
    	\begin{minipage}{0.29\columnwidth}
    		\begin{minted}[fontsize=\scriptsize,escapeinside=&&,autogobble,highlightlines={2,3},highlightcolor=lYellow!20]{coq}
&$\mathcal{ST}$&: State,
&$\mathcal{Z}$&: (Challenge x Signature)
    		\end{minted}
    	\end{minipage}
    };

    \draw[] (code-i.north west) -- (code-i.north east);
    \draw[] (code-i.north west) |- (state-i.north) -| (code-i.north east);
    \draw[] (state-i.north west) |- (ideal.north) -| (code-i.north east);
    \draw[] (code-i.north east) |- (code-i.south) -| (code-i.north west);

\end{tikzpicture}
  \caption{Game for \ra Primitives}
  \label{fig:att:prim:game}
\end{figure}
\vspace{-0.5em}
We first establish perfect indistinguishability between \ra primitives 
and their lifted signature counterparts.
\vspace{-0.4em}
\begin{lemma}[Perfect IND for \real Attestation Primitives]\label{lem:prim:real}
    For every adversary \A, the advantage of distinguishing the \real packages for \ra primitives
    is $0$: 
    \begin{center}
        \begin{minipage}{0.5\columnwidth}
    \begin{minted}[fontsize=\footnotesize,escapeinside=&&,autogobble]{coq}
SigPrimAtt&$_{\text{\real}}$& &$\pindist$& AttPrimSig&$_{\text{\real}}$&.
    \end{minted}
    \end{minipage}
    \end{center}
\end{lemma}
\vspace{-0.4em}
\begin{proof}
    The proof is by reflexivity on the fact
    that \pattprim and \pattprim{$_{\real}$} are
    equal.
\end{proof}
\vspace{-0.4em}
\vspace{-0.4em}
\begin{lemma}[Perfect IND for \ideal Attestation Primitives]\label{lem:prim:ideal}
    For every adversary \A, the advantage of distinguishing  
    the \ideal packages for \ra primitives
    is $0$: 
    \begin{center}
        \begin{minipage}{0.5\columnwidth}
    \begin{minted}[fontsize=\footnotesize,escapeinside=&&,autogobble]{coq}
SigPrimAtt&$_{\text{\ideal}}$& &$\pindist$& AttPrimSig&$_{\text{\ideal}}$&.
    \end{minted}
    \end{minipage}
    \end{center}
\end{lemma}
\vspace{-0.4em}
\vspace{-0.4em}
\begin{proof}
    Instead of discarding the new memory location
    $\mathcal{Z}$ in \pattprim{$_{\ideal}$}, our proof 
    connects it to the $\mathcal{SIG}$ 
    location in \psigprim{$_{\ideal}$} with the
    following invariant:
  \vspace{-0.5em}
  \begin{minted}[escapeinside=@@,fontsize=\footnotesize,autogobble]{Coq}
 @$\lambda$@ (s@$_1$@,s@$_2$@), heap_ignore @$\{\mathcal{Z}\}$@ (s@$_1$@,s@$_2$@) @$\land$@
            Z_to_SIG s@$_2$@.@$\mathcal{Z}$@ s@$_2$@.@$\mathcal{ST}$@ = s@$_1$@.@$\mathcal{SIG}$@.
\end{minted}
\vspace{-0.5em}
    The \hignore invariant defines equality on
    all state location except $\mathcal{Z}$.
    The function \ztosig translates $\mathcal{Z}$
    to $\mathcal{SIG}$ via the hash function $\mathcal{H}$.
\vspace{-0.5em}
\begin{minted}[escapeinside=@@,fontsize=\footnotesize,autogobble]{Coq}
Definition h_to_sig @$\mathcal{Z}$@ @$\mathcal{ST}$@:
  let s := @$\mathcal{ST}$@ in
  map (@$\lambda$@ (c,a), (@$\mathcal{H}$@ c s, a)) @$\mathcal{Z}$@.
\end{minted}
\vspace{-0.5em}
    The proof consists of three proof obligations, one per procedure.
    The details are in our proof development.
    The most interesting part is in the proof obligation
    for \everifya, where we have to establish the
    following equality:
\vspace{-0.5em}
\begin{minted}[escapeinside=@@,fontsize=\footnotesize,autogobble]{Coq}
(@$\mathcal{H}$@ c@$_1$@ s@$_1$@, a) @$\in$@ @$\mathcal{SIG}$@ = (c, a) @$\in$@ @$\mathcal{Z}$@ 
\end{minted}
\vspace{-0.5em}
    By applying our invariant and rewriting the left-hand side, we derive:
\vspace{-0.5em}
\begin{minted}[escapeinside=@@,fontsize=\footnotesize,autogobble]{Coq}
(@$\mathcal{H}$@ c@$_1$@ s@$_1$@, a) = (@$\mathcal{H}$@ c@$_2$@ s@$_2$@, a) 
\end{minted}
\vspace{-0.5em}
    Note that the challenges and the states do not
    unify immediately because c$_2$ and s$_2$ come from
    the invariant.
    But unification only arises in the relational Hoare logic
    reasoning.
    To solve this goal, we need a vital property of the hash
    function $\mathcal{H}$; collision-resistance, a.k.a., injectivity.
\end{proof}
\vspace{-0.5em}
\vspace{-0.5em}
\begin{hypothesis}[Collision-Resistance]
    Any hash function $\mathcal{H}$ used in \ra
    must be injective:
    \vspace{-0.5em}
    \begin{center}
    \begin{minipage}{0.7\columnwidth}
\begin{minted}[escapeinside=@@,fontsize=\footnotesize,autogobble]{Coq}
@$\forall$@ c@$_1$@ s@$_1$@ c@$_2$@ s@$_2$@, @$\mathcal{H}$@ c@$_1$@ s@$_1$@ = @$\mathcal{H}$@ c@$_2$@ s@$_2$@ -> c@$_1$@ = c@$_2$@ @$\land$@ s@$_1$@ = s@$_2$@.
\end{minted}
    \end{minipage}
    \end{center}
  \end{hypothesis}
\vspace{-0.5em}
\vspace{-0.5em}
\begin{theorem}[
    Security reduction for \ra primtives]\label{theo:redprim}
  For every adversary \A, the advantage of distinguishing 
  the packages \pattprim{$_{\text{\real}}$} and \pattprim{$_{\text{\ideal}}$} is
  less than or equal to the advantage of distinguishing 
  the packages: \psigprim{$_{\text{\real}}$} and \psigprim{$_{\text{\ideal}}$}
  \begin{center}
      \begin{minipage}{0.8\columnwidth}
     \begin{minted}[fontsize=\footnotesize,escapeinside=&&,autogobble]{coq}
  &$\forall$& A,&\spacer{0.1cm}& Adv A             AttPrim&$_{\text{\real}}$& AttPrim&$_{\text{\ideal}}$& &$\leq$& &\spacer{0.1cm}&
       Adv (A &$\circ$& AttPrim) SigPrim&$_{\text{\ideal}}$& SigPrim&$_{\text{\real}}$&.
    \end{minted}
     \end{minipage}
  \end{center}
\end{theorem}
\vspace{-0.5em}
\vspace{-0.5em}
\begin{proof}
    The proof follows the same structure as the reduction proof for the 
    protocols (Theorem~\ref{theo:redprot}):
    We use the triangle inequality and apply Lemmas~\ref{lem:prim:real} and \ref{lem:prim:ideal} afterward.
    %
    %
\end{proof}
\vspace{-0.6em}

  	\section{Related Work}\label{sec:rel}
\vspace{-0.5em}
Early work by Cabodi et al.~\cite{cabodi2015formal, cabodi2016secure} laid the foundation 
for formalizing hybrid \ra properties, 
analyzing and comparing \ra architectures using different model checkers. 
VRASED~\cite{vrased} extended this by introducing 
a verified hardware-software co-design for \ra, targeting low-end embedded systems 
and addressing security limitations of hybrid architectures~\cite{literature, minimalist}.
Hybrid \ra mechanisms, such as HYDRA~\cite{hydra}, further enhanced security by 
integrating formally verified microkernels, achieving memory isolation 
while reducing hardware complexity.
Formal verification has also been applied 
to the industry~\cite{sardar2021demystifying} 
for one of Intel TDX's security-critical processes using ProVerif. 
Most symbolic approaches (e.g., ProVerif~\cite{proverif12} and Tamarin~\cite{tamarin}) are used to 
model attacks~\cite{de2021toctou, seed17} but offer limited support for 
computational reductions or indistinguishability arguments. 
While these tools effectively validate protocol design~\cite{wesemeyer2020formal}, 
they cannot capture strong cryptographic assumptions such as unforgeability 
and negligible advantage. 
From a cryptographic proof perspective, tools like EasyCrypt~\cite{barthe2012easycrypt} 
support formal game-based security proofs~\cite{shoup2004sequences}. 
Recent work in EasyCrypt has been employed to verify existential~\cite{dupressoir2021machine}
and strong unforgeability~\cite{cortier2018machine}, 
but applying these techniques to \ra has not been explored. 
Our work is the first step toward fully machine-checked, game-based security proofs of \ra, 
formalizing semantic security through perfect indistinguishability in the computational model.

	\section{Conclusion}\label{sec:concl}
\vspace{-0.6em}
We presented a formal verification of semantic security for digital signatures 
and \ra using the \ssprove framework in the \coq. 
To the best of our knowledge, this is the first work to establish 
perfect indistinguishability in a machine-checked, 
game-based setting for \ra protocols. 
This work formalizes and proves security against forgery in \ra, 
providing a foundational guarantee based on sEUF-CMA. 
Our development reduces the security of \ra to the security of secure signatures. 
We showed that the perfect indistinguishability of 
signatures can be lifted to prove the security of \ra protocols. 

In future work, we aim to extend our formalization to verify 
complex security protocols~\cite{fixedtpm17} and 
strengthen \ra{’s} overall security~\cite{de2021toctou,seed17,forkreply07}.
%
Beyond SSProve, we intend to integrate our formalization 
into the Clutch framework, which supports 
probabilistic relational reasoning in Iris~\cite{jungiris15} 
via ghost state and presampling tapes and may allow 
formalizing attestation semantics in more expressive systems~\cite{asynchronous24}.
%
%
%
\vspace{-0.6em}
\section{Acknowledgements}
We appreciate Dr. Carmine Abate and the anonymous reviewers for their valuable feedback. 
This work has been supported by the Federal Ministry of Research, Technology and 
Space of Germany through the 6G-ICAS4Mobility project (grant no. 16KISK231) and 
the 6G-Plattform Germany project (grant no. 16KISK046). 
Additionally, the authors are also financed based on the budget passed by 
the Saxonian State Parliament in Germany.
\vspace{-0.6em}


	\bibliographystyle{splncs04}
	\bibliography{main}

    \section{Appendix}
    \paragraph{\textbf{RSA-based Digital Signatures}}
\label{sec:rsa}
\vspace{-0.6em}
\vspace{-0.5em}
\begin{figure}
    \centering
    \begin{tikzpicture}[
	scale=0.75, transform shape
	]
	\node[] (P) at (0,0) {
		\begin{minipage}{0.28\columnwidth}
			\begin{minted}[linenos,fontsize=\scriptsize,escapeinside=&&,autogobble,highlightlines={2,3,8},highlightcolor=lGray!20]{coq}
def RSA.KeyGen ():
  p <$ P ;;
  q <$ Q p ;;
  assert (p != q) ;;

  let &$\phi^n$& := (p-1)*(q-1) in

  e <$ E &$\phi^n$& ;;
  let d := e&$^{-\texttt{1}}$& (mod &$\phi^n$&) in
  let ed := e * d in
  assert (ed == 1) ;;

  let n := p * q in
  let pk := (n,e) in
  let sk := (n,d) in

  #ret (pk,sk)
			\end{minted}
		\end{minipage}
	};
	\node[] (s) [above right = 0cm and 0cm of c.north west] {
    \begin{minipage}{0.65\columnwidth}
		\begin{minted}[fontsize=\scriptsize,escapeinside=&&,autogobble]{coq}
      &$\phantom{empty-line}$&
        \end{minted}
    \end{minipage}
	};

\end{tikzpicture}
    \caption{RSA-based Key Generation}
    \label{fig:rsa:keygen}
\end{figure}
\vspace{-0.5em}
To make sure that our Hypothesis~\ref{def:sig:correct}
for functional correctness of digital signatures as
stated in Figure~\ref{hypo:funCorssprove} is indeed
sufficient, we implement RSA-based digital signatures.
RSA-based signatures are one of the schemes from the
TPM 2.0 specification.
We take full advantage of \ssprove and its access to the
rich ecosystem of existing developments.
A full definition of RSA along with a proof of
functional correctness is available in the \texttt{mathcomp-extra}
library.\footnote{https://github.com/thery/mathcomp-extra}
We placed this proof at the heart of our proof
for functional correctness of RSA-based digital signatures.
A substantial part of our development then evolves around
the setup of the sampling spaces.
In Figure~\ref{fig:rsa:keygen}, we list a condensed and
slightly simplified versioni of the code for 
$\Sigma$.\icoq{KeyGen}.
The interested reader can find the full implementation
in our \coq development.
\vspace{-0.5em}
\paragraph{\textbf{Sample Spaces.}}
The code highlights the three places where we sample
values from uniform distributions.
In total, we need to sample values for $p$, $q$ and $e$
(Lines~2,3 and 8).
From these values, we calculate $d$ (Line~9) and
define the public and the secret key to return (Lines~13--17).
Our sampling spaces $P$, $Q$ and $E$ establish the
properties that are need for our functional correctness
proof.
The space that all threee spaces are based upon must
obey the following requirements.
First, we need to sample prime numbers.
Second, there need to exist at least three distinct prime
numbers.
The following is our mathcomp-based definition:
\vspace{-0.5em}
\begin{minted}[fontsize=\scriptsize,escapeinside=&&]{coq}
Variable n : &$\mathbb{N}$&.
Definition &$\mathbb{B}$& : Type := 'I_(n.+6).
Definition primes : finType := {x :&$\mathbb{B}$& | prime x}.
Definition P : {set primes} := [set : primes].
\end{minted}
\vspace{-0.5em}
This definition fulfills both requirement.
The set \icoq{P} contains only primes.
And \icoq{P} contains at least the primes that are smaller 
than $6$, i.e., $2$, $3$ and $5$.
The sampling space \icoq{Q} now needs to establish the
property that \icoq{p != q} (Line~4).
That is, \icoq{Q} is depending on the value that was 
sampled before from \icoq{P}.
But \icoq{Q} has to establish yet another property for
the space \icoq{E}:
\vspace{-0.5em}
\begin{minted}[fontsize=\scriptsize,escapeinside=&&]{coq}
Definition Q p :=
  let P' := P :\ p in
  if p == 2 
  then P' \ 3
  else if p == 3 
       then P' :\ 2
       else P'.
\end{minted}
\vspace{-0.5em}
In \icoq{RSA.KeyGen}, we use \icoq{p} and \icoq{q}
to compute $\phi^n$, the Euler totient function (Line~6).
This $\phi^n$ then defines $E$:
\vspace{-0.5em}
\begin{minted}[fontsize=\scriptsize,escapeinside=@@]{coq}
Definition E' {m: @$\mathbb{B}$@ * @$\mathbb{B}$@)} (H: 2<m) :=
  { x : @$\mathbb{Z}_m$@ | 1 < x && coprime m x}.
Definition E {m: @$\mathbb{B}$@ * @$\mathbb{B}$@)} (H: 2<m) : {set (E' H)} := 
  [set : E' H].
\end{minted}
\vspace{-0.5em}
$E$ needs to have at least one value to sample \icoq{e} (Line~8).
That is, $\phi^n > 3$.
And hence, the construction of $Q$ must exclude the
two cases where \icoq{p := 2} and \icoq{q := 3} or vice versa
because 
\vspace{-0.6em}
\[
    \phi^n = (2-1) * (3-1) = 5 \ngtr 5
\]
\vspace{-0.6em}
would not provide a space to sample a coprime number \icoq{e} from.
\subsection{Functional Correctness}
\vspace{-0.5em}
Based on this sample space construction, we can now
proof functional correctness for our digital signature scheme.
\vspace{-0.5em}
\begin{theorem}[Functional Correctness for RSA-based Digital Signatures]
    Given the definitions of our RSA-based signature scheme $\Sigma := \texttt{RSA}$,
    the following holds:
  \begin{center}
    \begin{minipage}{0.75\columnwidth}
    \begin{minted}[fontsize=\scriptsize,escapeinside=@@,autogobble]{coq}
@$\forall$@ m sk pk,
    ((sk,pk) <- RSA.KeyGen) ->
    RSA.VerSig pk (RSA.Sign sk m) m == true.
    \end{minted}
    \end{minipage}
    \end{center}
\end{theorem}
\vspace{-0.5em}
\begin{IEEEproof}
    The proof establishes all necessary properties
    from the construction of the sampling spaces and
    finally reduces to the functional correctness
    of RSA itself.
    Our \coq development has all the details.
\end{IEEEproof}
\vspace{-0.5em}

\end{document}